\documentclass[12pt]{iopart}

\usepackage{iopams}  

\usepackage{etoolbox}
\makeatletter
\def\@mkboth#1#2{}
\newlength\appendixwidth
\preto\appendix{\addtocontents{toc}{\protect\patchl@section}}
\newcommand{\patchl@section}{%
	\settowidth{\appendixwidth}{\textbf{Appendix }}%
	\addtolength{\appendixwidth}{1.5em}%
	\patchcmd{\l@section}{1.5em}{\appendixwidth}{}{\ddt}%
}
\makeatother

\expandafter\let\csname equation*\endcsname\relax
\expandafter\let\csname endequation*\endcsname\relax
\usepackage{amsmath}

\usepackage{mathtools}

\usepackage{mathrsfs}

\usepackage{graphicx}

\usepackage[dvipsnames]{xcolor}
\usepackage[colorlinks,linkcolor=blue,citecolor=blue]{hyperref}

\usepackage{tikz}
\usetikzlibrary{arrows.meta}
\usepackage{pgfplots}
\pgfplotsset{width=7cm,compat=1.8}
\usepgfplotslibrary{fillbetween}

\usepackage{enumerate}
\usepackage{cite}
\usepackage{empheq}
\usepackage{placeins}
\usepackage{float}
\usepackage{mathrsfs}
\usepackage{adjustbox}
\usepackage{mwe}
\usepackage{cases}

\allowdisplaybreaks

\newcommand*\diff{\mathop{}\!\mathrm{d}}

\renewcommand{\vec}[1]{\boldsymbol{#1}}
\newcommand{\del}{\partial}

\newcommand{\avg}[1]{\left\langle#1\right\rangle}

\newcommand{\vw}{\vec{w}}
\newcommand{\vx}{\vec{x}}

\newcommand{\vv}{\vec{v}}
\newcommand{\vu}{\vec{u}}

\newcommand{\vq}{\vec{q}}

\newcommand{\vtau}{\vec{\tau}}

\begin{document}
	
	\title[Inter-particle ratchet effect determines current of heterogeneous confined particles]{Inter-particle ratchet effect determines global current of heterogeneous particles diffusing in confinement}
	
	\author{E Mallmin$^1$, Richard A. Blythe$^1$, and Martin R. Evans$^1$}
	
	\address{$^1$ SUPA, School of Physics and Astronomy, University of Edinburgh, Peter Guthrie Tait Road, Edinburgh EH9 3FD, UK}
	\ead{emil.mallmin@ed.ac.uk}
	\vspace{10pt}
	\begin{indented}
		\item[]December 2020
	\end{indented}
	
	\begin{abstract}
		In a model of $N$ volume-excluding spheres in a $d$-dimensional tube, we consider how differences between particles in their drift velocities, diffusivities, and sizes influence the steady state distribution and axial particle current. We show that the model is exactly solvable when the geometrical constraints prevent any particle from overtaking every other---a notion we term quasi-one-dimensionality. Then, due to a ratchet effect, the current is biased towards the velocities of the least diffusive particles. We consider special cases of this model in one dimension, and derive  the exact joint gap distribution for  driven tracers in a passive bath. We describe the relationship between phase space structure and irreversible drift that makes the quasi-one-dimensional supposition key to the model's solvability.
	\end{abstract}
	
	\begin{indented}
		\item[\sffamily\bfseries Keywords. ] Single-file diffusion, stochastic ratchet, nonequilibrium steady states 
	\end{indented}
	
	\vspace{2cm}
	\noindent Published in JSTAT\\
	\noindent DOI: \href{https://doi.org/10.1088/1742-5468/abd30e}{10.1088/1742-5468/abd30e}
	
	%
	%
	%
	%
	%

	\maketitle
	
	\newpage

\noindent\rule{\textwidth}{2pt}
\tableofcontents

\noindent\rule{\textwidth}{2pt}

\FloatBarrier

\section{Introduction and summary}

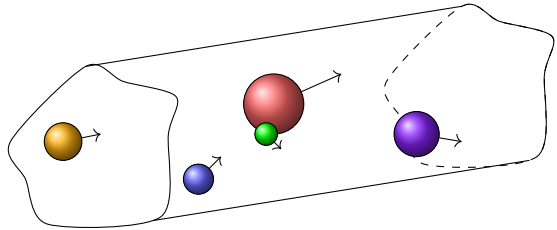
\begin{figure}[t]
	\begin{center}		
		\begin{tikzpicture}
		\def \y {0.8};
		\def \x {5};
		
		\draw [-] plot [smooth cycle] coordinates {(0,0) (1,1) (1.5,0.8) (2.2,0.6) (2.1,0.1) (2,-1)  (0.5,-1.1) (0.2,-0.5)};
		
		\draw [dashed] plot [smooth cycle] coordinates {(0+\x,0+\y) (1+\x,1+\y) (1.5+\x,0.8+\y) (2.2+\x,0.6+\y) (2.1+\x,0.1+\y) (2+\x,-1+\y)  (0.5+\x,-1.1+\y) (0.2+\x,-0.5+\y)};
		
		\begin{scope}
		\clip (1+\x,1+\y) -- (2+\x,-1+\y) -- (2.5+\x,-1+\y) -- (2.5+\x,1.8+\y) -- cycle;
		\draw [-] plot [smooth cycle] coordinates {(0+\x,0+\y) (1+\x,1+\y) (1.5+\x,0.8+\y) (2.2+\x,0.6+\y) (2.1+\x,0.1+\y) (2+\x,-1+\y)  (0.5+\x,-1.1+\y) (0.2+\x,-0.5+\y)};
		\end{scope}	
		
		\draw[-] (1,1)--(1+\x,1+\y);
		\draw[-] (1.9,-1.05)--(1.9+\x,-1.05+\y);		
		
		\draw[->] (0.7,0)--(1.2,+0.1);
		\shade [ball color=Yellow!50!Orange] (0.7,0) circle (0.25);
		\draw (0.7,0) circle (0.25);
		
		\draw[->] (2.5,-0.5)--(2.8,-0.2);
		\shade [ball color=Blue!60!White] (2.5,-0.5) circle (0.2);
		\draw (2.5,-0.5) circle (0.2);
		
		\draw[->] (3.5,0.5)--(4.4,0.9);
		\shade [ball color=Red!60!White] (3.5,0.5) circle (0.4);
		\draw (3.5,0.5) circle (0.4);
		
		\draw[->] (3.4,0.1) -- (3.6,-0.1);
		\shade [ball color=Green] (3.4,0.1) circle (0.15);
		\draw (3.4,0.1) circle (0.15);
		
		\draw[->] (5.4,0.1)--(6,0);
		\shade [ball color=Purple] (5.4,0.1) circle (0.3);
		\draw (5.4,0.1) circle (0.3);
		
		\end{tikzpicture}
	\end{center}
	\caption{Tube geometry: constant cross section and periodic or infinite axial direction. Illustrated here for dimension $d = 3$.}\label{fig:tube}
\end{figure}

Models of colloidal particles diffusing in narrow channels constitute an important class of nonequilibrium systems, being both empirically relevant and readily analysed analytically or \emph{in silico}\cite{Burada2009}. A principal feature of such systems is that directed particle transport generically arises as out-of-equilibrium conditions conspire with spatial, temporal, or noise-induced asymmetries to rectify fluctuations\cite{Hanggi2009,Cubero2016}. Iconic studies of a single particle ratcheted by an asymmetric potential \cite{Magnasco1993,Doering1994} have been complemented by descriptions of inherently many-body ratchet phenomena, e.g.\ collectively induced asymmetric ratchet potentials in magnetic vortices in superconducting films \cite{Olson2001}, density-dependent current reversals \cite{Derenyi1995,SouzaSilva2006}, and active matter motion rectified by asymmetric obstacles \cite{OlsonReichhardt2017}. In this article, we demonstrate how even in a static, structureless environment, particles can themselves be the source of inter-particle ratcheting when they have heterogeneous properties and interact via volume exclusion. This inter-particle ratchet effect underlies correlation-phenomena found in a range of models including  lattice exclusion processes with particle-wise disorder \cite{Evans1996,Krug1996,Evans1997} or driven tracers \cite{Lobaskin2020,Miron2020}, continuous-space single-file diffusion with random diffusivities \cite{Aslangul2000,Ambjornsson2008,Flomenbom2010} or friction \cite{Lomholt2014}, and a recent many-filament ratchet model \cite{Wood2019}. 

To demonstrate the inter-particle ratchet effect analytically, we consider the steady-state properties of $N$ hard $d$-spheres in a $d$-dimensional periodic `tube' (\autoref{fig:tube}). In particular, we are interested in the net particle current and how it is shaped by collective effects arising from the interplay of particle heterogeneity and interactions. Heterogeneity means that the particles can have different sizes, and experience differently the (time-homogeneous) external or internal drive, and thermal noise. In other words, there is a quenched disorder in the particle properties. Interactions occur between particles and with the tube boundaries via volume exclusion. A key insight of this work is that the model is solvable when all particles have a \textit{common net velocity}, defined as the long-time net translation per unit time given the presence of  interactions. (This is equivalent to the net current up to a choice of units.) For instance, the requirement of common net velocity holds in the common scenario where the tube is narrow enough that particles cannot overtake at all. But it also holds when particles \emph{can} overtake, as long as no particle can overtake \emph{all} others, or when there is a movable barrier which prevents the particles from dispersing. We will adapt existing terminology and call either situation \emph{quasi-one-dimensional} (q1D). In the q1D case with spatially constant drift and diffusion coefficients, we derive the exact expression for the current. This reveals the ratchet effect, by which the least diffusive particles, i.e.\ the most deterministic, exert the greatest influence on the current. Furthermore, as far as the current is concerned, most model features---the tube dimension $d$, the shape of the (constant) tube cross section, whether the tube is periodic or infinite, the relative sizes of all particles, and even which pairs of particles are mutually volume-excluding---are not directly relevant. This is evidence for the robustness of the effect.

In \autoref{sec:model} below, we give the mathematical definition of the model as introduced above, and define the notation necessary to express our main result, Eq.\ \eqref{eq:intro-nu}, for the common net velocity $\tilde{v}$. The bulk of this work is then structured into two complementary parts which can be read independently:

In \autoref{sec:ex} we consider three one-dimensional special cases of the general model, presenting novel detailed calculations and interpretations of the result in terms of a ratchet effect. First, we study single-file diffusion on a ring with fully heterogeneous particle properties and derive Eq.\ \eqref{eq:intro-nu} for $d=1$. If the inverse diffusivities are drawn from a heavy-tailed distribution, corresponding, e.g., to a Gamma distribution for the diffusivities, one particle will dominate the current due to the ratchet effect. The particle heterogeneity, conceived of as a quenched disorder, is then not self-averaging. Secondly, by specializing to only two distinct particle species, we can model driven tracers in a passive medium. The tracers cooperate through ratcheting to achieve a current that scales linearly with the number of tracers. Lastly, we describe how the many-filament ratchet of Ref.\ \cite{Wood2019} relates to the previous cases through a simple mapping.

In \autoref{sec:defsol} we take a more formal approach in order to understand when and why the general tube model affords a solution through direct integration of the steady-state equation---this is the meaning we assign to the term `integrable' in this work. To this end, we develop an inverted method of finding solutions to stationary Fokker-Planck equations. Instead of solving directly for the density, one attempts to solve for the `irreversible drift' \cite{Graham1971}. By making an ansatz on the latter which is consistent with the boundary conditions, one implicitly introduces constraints on the model parameters which are in essence integrability conditions. In the tube model, we show that an ansatz of constant direction for the irreversible drift is consistent when the process is q1D, but otherwise not. The integrability conditions tell us which interaction potentials, beyond the volume exclusion, can be added between the particles without breaking the solution structure. In the absence of any such potentials it transpires that while the model is decidedly out-of-equilibrium, there exists a moving reference frame relative to which the system obeys detailed balance---this explains the solvability. We must emphasize, however, that this is not a trivial situation of an ``equilibrium system on wheels'', because the common net velocity emerges microscopically from the interaction of particles through a non-equilibrium effect, rather than being dictated by fiat. 

In the concluding discussion, \autoref{sec:disc}, we summarise  our findings and discuss them from the point of view of non-dissipative effects in nonequilibrium systems.

\section{Model definition and main result}\label{sec:model}

Let us now present the model and main results in more detail. The spherical particles exist in a tube which has one periodic axial direction $\hat{\vec{r}}$ (although we will show how an infinite tube can be treated as well), and a cross section that is constant along the tube, but otherwise of arbitrary shape---our focus is on inter-particle effects rather than particle-boundary effects, e.g.\ stemming from periodically varying channels \cite{Malgaretti2012,Malgaretti2017}. We consider overdamped motion, so that the system is fully described by the spatial coordinates $\mathbf{Q}^{(i)}(t)$ in $d$ dimensions for the particles $i=1 ,\ldots, N$. We organize these into
\begin{equation}\label{eq:X}
\mathbf{X}(t) = \begin{pmatrix}
\mathbf{Q}^{(1)}(t) \\
\vdots\\
\mathbf{Q}^{(N)}(t)\\
\end{pmatrix}
= \sum_{i=1}^N \hat{\vec{e}}_i \otimes
\mathbf{Q}^{(i)}(t),
\end{equation}
where $\hat{\vec{e}}_i$ is a standard basis vector of $\mathbb{R}^N$, $\otimes$ is the Kronecker product \cite{KroneckerProduct}. $\mathbf{X}(t)$ is thus an $N\times d$-dimensional drift-diffusion process evolving in a state space $\Gamma$ representing all possible configurations of the particles inside the tube geometry (the explicit construction of $\Gamma$ is given in \autoref{sec:defs}). We define the model dynamics through the formal stochastic differential equation
\begin{equation}\label{eq:sde}
\diff \mathbf{X} (t) = \vv \diff t +  \mathsf{B} \diff \mathbf{W}(t) + \diff \mathbf{R}(t).
\end{equation}
Here, $\vv$ is the spatially constant drift vector, $\mathsf{B}$ a spatially constant noise amplitude matrix, $\diff \mathbf{W}$ white noise, and $\diff \mathbf{R}$ a reflection term. The drift vector can be split as
\begin{equation}\label{eq:v}
\vv = \sum_{i=1}^N \hat{\vec{e}}_i \otimes \vv^{(i)},
\end{equation}
and $\vv^{(i)}$ interpreted as either an external force applied uniquely to particle $i$ (e.g.\ due to its unique charge $q_i$), or as an intrinsic velocity generated by it through self-propulsion. We define the diffusion matrix $\mathsf{D} = (1/2)\mathsf{B}\mathsf{B}^\top$. In the simplest scenario, $\mathsf{D}$ is diagonal with non-zero elements $D_i$ that may for instance be specified via the Einstein relation $D_i = \mu_i k_\text{B} T $ with $\mu_i$ the unique, size-dependent mobility of particle $i$. Alternatively, $D_i$ may relate to fluctuations in the self-propulsion mechanism under the active-matter interpretation of the model. However, we allow more generally for $\mathsf{D}$ to be non-diagonal, with spatially anisotropic diffusion and noise correlations between particles. The reflection term $\diff\mathbf{R}$ represents particle--particle and particle--wall contact forces and is only `active' when $\mathbf{X}(t) \in \del \Gamma$, the phase space boundary. It then simply reflects the process back into the interior of $\Gamma$. (This is the ``boundary local time'' implementation of reflection \cite{Grebenkov2019}, but we will instead unambiguously define reflection in the density-picture below.)

The process approaches a steady state whose density is defined 
\begin{equation}\label{eq:P(x)}
P(\vx) = \lim_{t\to\infty} \text{Prob}(\mathbf{X}(t) = \vx\,|\,\mathbf{X}(0) = \vec{x}_0).
\end{equation}
In the interior of the phase space $\Gamma$, the stationarity condition is 
\begin{equation}
\nabla  \cdot \vec{J}(\vx) = 0,
\end{equation}
using the probability current 
\begin{equation}\label{eq:J}
\vec{J}(\vx) = \vv  P(\vx) - \mathsf{D} \nabla P(\vx).
\end{equation}
The volume exclusion appears as reflective boundary conditions on  $\del\Gamma$, meaning that the probability current normal to the boundary vanishes:
\begin{equation}\label{eq:J.n-intro}
\vec{J}(\vx) \cdot \hat{\vec{n}}(\vx) = 0,\quad \vx \in\del \Gamma.
\end{equation}
Periodicity conditions on $P(\vx)$ make it invariant under simultaneously translating all particles by one axial period $L$,
\begin{equation}\label{eq:period}
P(\vx) = P(\vx + L \vtau),
\end{equation}
where the translation vector $\vtau$ is the direction in phase space of moving all particles forward in the axial direction $\hat{\vec{r}}$ by one unit length: 
\begin{equation}
\vtau = \sum_i \hat{\vec{e}}_i \otimes  \hat{\vec r}.
\end{equation}
(Note that $\vtau$ is not a vector in $\Gamma$ but belongs to its tangent space at any point.) If any particle is able to overtake all others, i.e.\ the system is not q1D, then $P(\vx)$ must be invariant under translating that particle alone by one period. This additional condition  
would break an important symmetry that allows the steady state to be determined, and so we concentrate on q1D where this condition does not hold. 

Our main result for this model is that
under q1D conditions a common net velocity $\tilde{v}$ emerges, which is  equal for all particles and is obtained from the probability current via integration as
\begin{equation}\label{eq:intro-nu}
\tilde{v} = \frac{1}{N} \int d\vx\, \vtau \cdot \vec{J}(\vx) = \frac{\vtau^\top \mathsf{D}^{-1} \vv}{\vtau^\top \mathsf{D}^{-1} \vtau}.
\end{equation}
\hypertarget{foo}{In} the trivial case where all particles diffuse (i) independently, (ii) isotropically, and (iii) identically, the common net velocity is simply 
\begin{equation}
\tilde{v} = \hat{\vec{r}} \cdot \frac{1}{N} \sum_i \vv^{(i)},
\end{equation}
as for unconfined, non-interacting particles. We are interested in violations of these conditions where the common net velocity is nontrivial. Violations of the three conditions are related to the structure of the matrix $\mathsf{D}$ which in different ways biases the current. The examples considered in the next section all relate to violating condition (iii); the other ones we touch upon in the discussion \autoref{sec:disc}.

\section{One-dimensional special cases}\label{sec:ex}

\subsection{Single-file diffusion heterogeneous particles}\label{sec:SFD}

\subsubsection{Introduction}
In one spatial  dimension, heterogeneous volume-excluding particles have been studied as minimal models of single-file traffic both on- \cite{Evans1996,Krug1996,Evans1997,Goncalves2008} and off-lattice \cite{Aslangul2000,Ambjornsson2008,Flomenbom2010,Lomholt2011,Lomholt2014}. In the lattice setting, the problem drew inspiration from vehicular or motor-protein transport. A drift-diffusion version of the problem can be physically motivated by noting that colloidal particles differing in size and electrical charge will have different thermal diffusivities (as given by the Einstein relation) and drift velocities in response to an applied electrical field. 

On the lattice, the model is the (totally) asymmetric exclusion process ((T)ASEP) with  particle-dependent hopping rates. For periodic boundary conditions its steady state was solved by a mapping to the zero-range process with site-wise disorder \cite{Evans1996,Krug1996,Evans1997,Lobaskin2020}. In the ASEP, each particle $i$ out of $N$ has a forward (clockwise) hopping rate $p_i$ and a backwards rate $q_i$ (with all $q_i = 0$ for TASEP). It was shown for the TASEP that if the $p_i$ are drawn independently from a power-law $f(p) \sim (p - c)^\gamma$ with support in $[c,1]$,  platoon-formation---$O(N)$ queue-formation behind the slowest particle---occurs above a critical density of walkers if $\gamma>0$, in a way mathematically analogous to Bose-Einstein condensation. While the more general ASEP was also solved exactly for any realization of the disorder, its behaviour as a function of different disorder distributions was difficult to analyze due to the complicated dependence of the steady-state on the model parameters \cite{Evans1996,Evans1997}. In the continuum, diffusive model studied in the present work we shall see that we do not observe the Bose-Einstein like condensation behind the slowest particles; instead  it is the diffusivities that control the common net velocity. 

On the continuum, the analogue of a one-dimensional exclusion process  is conventionally referred to as single-file diffusion (SFD) \cite{Ryabov2015} (although some make no distinction in terminology on/off-lattice). SFD has been previously studied for disordered diffusivities and zero drift \cite{Aslangul2000,Ambjornsson2008}, zero diffusivity and disordered drift (i.e.\ deterministic motion) \cite{BenNaim1994}, or identical diffusivity but disordered drift \cite{Lomholt2011,Lomholt2014}. The focus in these studies is on density relaxation and the scaling of the mean square displacement of a tagged particle in the long-time limit, which is famously sub-diffusive \cite{Harris1965}. In contrast, our focus is on the steady-state density and current, which would be trivial unless drifts are heterogeneous. Let us also point out that so far we know of no exact time-dependent solution to the problem of simultaneously heterogeneous diffusivities and drifts; critically, the Bethe ansatz that solves the identical-particle SFD \cite{Rodenbeck1998,Lizana2008,Lizana2009, Ryabov2015} fails in the presence of such heterogeneity as a crucial particle-exchange symmetry of the ansatz is violated. 

The main result of this section is the derivation of the steady-state density and current, and the expounding of the inter-particle ratchet effect that lets the least diffusive particle exert the greatest influence on the global current.

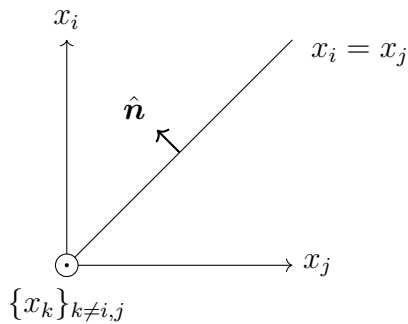
\begin{figure}
	\begin{center}		
		\begin{tikzpicture}
		\coordinate (Origin) at (0,0);
		\draw[->] (Origin)--(0,3) node[anchor=south] {$x_i$};
		\draw[->] (Origin)--(3,0) node[anchor=west] {$x_j$};
		\draw (Origin)--(3,3);
		\draw[fill=white] (0,0) circle (0.15) node[anchor=north,inner sep=3mm] {$\{ x_k \}_{k\neq i,j}$};
		\draw[fill=black] (Origin) circle (0.02);
		\node  at (3.9, 2.8)  {{$x_i = x_j$}};
		\draw[->,thick] (1.5,1.5)--(1.2,1.8) node[anchor=south east] {$\hat{\vec{n}}$};
		\end{tikzpicture}
	\end{center}
	\caption{Illustration of phase-space boundary}\label{fig:dSFDbc}
\end{figure}

\subsubsection{Exact solution}

The $N$ particles exist on a ring, each one $i$ moving with constant intrinsic drift velocity $v_i$ and experiencing thermal noise of amplitude  $\sqrt{2 D_i }$. All together, the positions, velocities and diffusivities are $\vx = (x_1,\ldots,x_N)^\top$, $\vv = (v_1,\ldots,v_N)^\top$, and $ \mathsf{D} = \text{diag}\{D_1,\ldots,D_N \}$. Two particles $i,j$ collide when $|x_i - x_j| = r_i + r_j$, with $r$ denoting the particle radius. As pointed out in \cite{Lizana2008} we may without loss of generality take $r_i = 0$ for all $i$. This is because what matters  is
 the free space between two  particles. The particle radii only
 enter through the total, conserved amount of free space throughout the system $L - \sum_{i=1}^N r_i$. The boundary conditions \eqref{eq:J.n-intro} prevent the  probability current $\vec{J}(\vx)$ \eqref{eq:J} from flowing past the  $x_i = x_j$ hypersurface in phase space, thereby preventing the particles from cross. From \autoref{fig:dSFDbc} one sees that $\hat{\vec{n}} = (\hat{\vec{e}_i} - \hat{\vec{e}_j})/\sqrt{2}$ and hence
\begin{equation}\label{eq:SFD-excl}
J_i(\vx,t) = J_j(\vx,t)\quad \text{for} \quad x_i = x_j.
\end{equation}

The steady state density $P(\vx)$ is solved for by an exponential ansatz, similar to how a factorized ansatz solves the lattice version of the model:
\begin{equation}\label{eq:SFD-P}
P(\vx) \propto e^{\vec{k} \cdot \vx} \times \text{\{ ordering constraint \}}.
\end{equation}
The ordering constraint means particles must appear in the same order on the ring as they do in the initial condition $\vx_0$. From the exclusion boundary condition \eqref{eq:SFD-excl} it follows that
\begin{equation}\label{eq:k_i}
k_i = \frac{v_i - \tilde{v}}{D_i},
\end{equation}
where $\tilde{v}$ (the common net velocity) is to be determined. 
 The periodicity condition \eqref{eq:period} requires that $\sum_i k_i = 0$, which implies that the common net velocity $\tilde{v}$ is given by
\begin{equation}\label{eq:SFD-nu-1}
\tilde{v} = \sum_{i=1}^N \left(  \frac{D_i^{-1}}{\sum_j D_j^{-1}} \right)  v_i .
\end{equation}
One can also check that \eqref{eq:SFD-nu-1}
satisfies the stationarity condition $\nabla \cdot \vec{J}(\vx) = 0$.

Introducing the \textit{effective diffusivity} $\widetilde{D}$ via 
\begin{equation}
\frac{1}{\widetilde{D}} = \sum_{i = 1}^N \frac{1}{D_i}
\end{equation}
we can express \eqref{eq:SFD-nu-1} more suggestively as
\begin{equation}
\frac{\tilde{v}}{\widetilde{D}} = \sum_{i=1}^N \frac{v_i}{D_i}.
\end{equation}
$ \widetilde{D}$ also features in the prefactor of the ($t\to \infty$) mean-square displacement of unbiased random walkers with heterogeneous diffusivities \cite{Ambjornsson2008,Goncalves2008}. Furthermore, it is formally equivalent to the formula for an effective spring constant replacing springs coupled in series.

To prove that $\tilde{v}$ is indeed the common net velocity, note that it is given by integrating the $i$th component $J_i(\vx)$ of the probability current over phase space, and presently    
\begin{equation}
J_i(\vx) = \tilde{v} P(\vx).
\end{equation}

The best strategy to determine the normalization constant of \eqref{eq:SFD-P} is to change variables to the `gaps' 
$y_i$ between a particle $i$ and its clockwise neighbour  particle $i+1$ (with $N+i$ equivalent to $i$),
\begin{equation}
y_i = x_{i+1} - x_i\;.
\end{equation}
The following notational convention is useful:
\begin{equation}\label{eq:sumnot}
y_{i:j} = y_i + y_{i+1} + y_{i+2} + \ldots + y_{j}= \begin{cases}
\sum_{l=i}^j y_l, &  i \leq j\\
0, & i > j
\end{cases}.
\end{equation}
We then go from absolute coordinates $x_i$ to gap coordinates via
\begin{equation}\label{eq:x->y}
x_i = x_1 + y_{1:i-1}.
\end{equation} 
This allows us to rewrite
\begin{equation}
\vec{k} \cdot \vx = x_1 \underbrace{\sum_{i=1}^{N} k_i}_{=0} + \sum_{i=1}^{N} k_i y_{1:i-1} = \sum_{i=1}^{N} k_{i+1:N}\, y_i,
\end{equation}
where the last equality follows from writing out the sums in the previous step term by term and summing up the coefficient for each $y_i$ before summing over $i$. For our convenience, let us define
\begin{equation}\label{eq:w_i}
w_i = k_{i+1:N} = - k_{1:i}.
\end{equation}
Then the exact  stationary probability distribution may be expressed as
\begin{equation}\label{eq:SFD-P-y}
P(\vx) = P_{\text{ctr}}(\vx) \times P_{\text{gaps}}(\vec{y}(\vx)) = \frac{1}{L} \times \frac{1}{ Z_N(\vw; L)} \exp[\vec{w}\cdot \vec{y}] \delta\left( \sum_{i=1}^{N} y_i - L \right),
\end{equation}
a product of the flat centre-of-mass distribution ($1/L$) and the gap distribution. 
Note that the gap distribution factorises into weights ${\rm e}^{w_i y_i}$ for each gap, but the  gaps are still correlated due to the global constraint that they sum to $L$, which is implied by the delta function.
The partition function $Z_N$ is defined
\begin{subequations}\label{eq:Z-def}
\begin{align}\label{eq:ZN=}
Z_N(w_1,\ldots,w_N; L) &= \left(\prod_{i=1}^N \int_0^L dy_i\, e^{ w_i y_i}\right)  \delta\left( L - y_{1:N} \right)\\\label{eq:ZNb}
& = \sum_{i=1}^{N} e^{w_i L} \prod_{j=1\neq i}^{N} \frac{1}{w_i - w_j}.
\end{align}
\end{subequations}
The integral in \eqref{eq:ZN=} can be shown to yield \eqref{eq:ZNb} by taking a Laplace transform in $L$, which decouples the integrals, thus allowing them to be computed. The inversion is then obtained using the Bromwich contour and the residue theorem.

\subsubsection{Demonstration of the ratchet effect}
We now consider the effect of the disorder in the particles' parameters  (velocities and diffusivities). We can write \eqref{eq:SFD-nu-1} as a weighted sum
\begin{align}
\tilde{v} &= \sum_i \alpha_i v_i,\label{eq:SFD-nu-weighted}\\
\alpha_i &= \frac{\widetilde{D}}{D_i}= \frac{D_i^{-1}}{\sum_j D_j^{-1}} \geq 0, 
\quad \sum_i \alpha_i = 1.
\end{align}
Now consider the $D_i$ to be i.i.d.\ random variables. The $\alpha_i$ are then \textit{dependent} but still identically distributed, which together with their conservation law implies $[\alpha_i]_D = 1/N$, where $[\cdot]_D$ denotes averaging over the diffusivity disorder. Hence
\begin{equation}
[\tilde{v}]_D = \frac{1}{N} \sum_{i=1}^N v_i,
\end{equation}  
the right-hand side of which is the sample mean over velocities, independently of the diffusivity distribution chosen. In particular, this is the same as for particles with identical diffusivity. If also the $v_i$ are considered i.i.d.\ random variables, with a well-defined mean $\bar{v}$, then
\begin{equation}
[\tilde{v}]_{D,v} = \bar{v}.
\end{equation} 

However, the \emph{typical} current $\tilde{v}^*$ may be very different from $\bar{v}$, meaning that the current is not self-averaging over the disorder. For this to be possible, there must be some spread in the velocities because if all $v_i = v$ then $\tilde{v} = v$ independently of the diffusivities. Now, \eqref{eq:SFD-nu-weighted} is a sum of i.i.d.\ random variables $R_i = \alpha_i v_i$. It is well known that if the distribution of $R$ has a power-law tail $\sim R^{-(a+1)}$ with $0< a < 1$, then the sum will be dominated by one particular $i$, say $i^*$  \cite{Derrida1993}. The physical significance of this is that particle  $i^*$ dictates the current. To illustrate a situation where self-averaging does not occur, assume finite support for the velocity distribution, and consider $\text{Prob}(1/D_i) \sim (D_i)^{a+1}$, $0 < a < 1$, for $D_i$ small, for example by drawing $D_i$ from the appropriate Gamma distribution. Then one $\alpha_{i^*}$ will be of order one and $\tilde{v} \approx v_{i^*}$. For instance, if the $v_i$ are $\pm |v|$ with equal chance, then the typical current will be either $\approx \pm |v|$, which is different from $[\tilde{v}]_{D,v} = 0$. 

Turning now to the distribution of particle gaps, ideally one would like to average the exact gap distribution \eqref{eq:SFD-P-y} over various disorder distributions. However, due to the complexity of e.g.\ the partition function \eqref{eq:Z-def}, the  average over the disorder is  forbiddingly difficult. Instead, we make progress by assuming a simple case where particle one, say, is dominating, so that $\tilde{v} \approx v_1$, whereas the other $N-1$ particles can approximately be treated as having identical parameters $v$ and $D$. Then 
\begin{equation}
\tilde{v} = \alpha v_1 + (1- \alpha)v,\quad \alpha =  \frac{1}{1 + (N-1) \frac{D_1}{D}},
\end{equation}
and for the non-dominant particles
\begin{equation}
k_i =  \frac{v - \tilde{v}}{D} = \frac{v - v_1}{D + (N-1)D_1}\equiv -k,
\end{equation}
whereas
\begin{equation}
k_1 = (N-1)k
\end{equation}
due to the conservation law $\sum_i k_i = 0$.

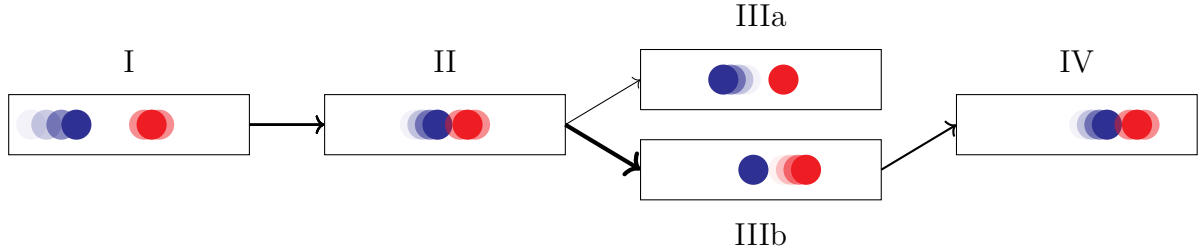
\begin{figure}
	\begin{tikzpicture}
	\def \L {3.2}
	\def \r {1}
	\def \h {0.4}
	\def \sep {0.7}
	
	\draw (0,-\h) rectangle ++(\L,\h*2) node[midway,above,inner sep=\sep cm] {I};
	
	\foreach[count=\i] \x in {0, 0.2, 0.4, 0.6}
	{
		\fill[Blue, opacity={\i*\i/16}] (0.3+\x, 0) circle (0.2);
	}
	\foreach[count=\i] \x in {0, 0.1, 0.2}
	{
		\fill[Red, opacity={1-(\i-2)*(\i-2)/2}] (1.8+\x, 0) circle (0.2);
	}
	\draw[->, thick] (\L,0) -++ (\r,0);
	
	\draw (\L+\r,-\h) rectangle ++(\L,\h*2) node[midway,above,inner sep=\sep cm] {II};
	\draw[->] (2*\L+\r,0) -++ (\r,1.5*\h);
	\draw[->,ultra thick] (2*\L+\r,0) -++ (\r,-1.5*\h); {IIb};
	\foreach[count=\i] \x in {0, 0.1, 0.2, 0.3}
	{
		\fill[Blue, opacity={\i*\i/16}] (\L+\r+\x+1.2, 0) circle (0.2);
	}
	\draw[->, thick] (\L,0) -++ (\r,0);
	\foreach[count=\i] \x in {0, 0.1, 0.2}
	{
		\fill[Red, opacity={1-(\i-2)*(\i-2)/2}] (\L+\r+1.8+\x, 0) circle (0.2);
	}
	\draw[->, thick] (\L,0) -++ (\r,0);

	\draw (2*\L+2*\r,1.5*\h-\h) rectangle ++(\L,\h*2) node[midway,above,inner sep=\sep cm] {IIIa};
	\foreach[count=\i] \x in {0, 0.1, 0.2, 0.3}
	{
		\fill[Blue, opacity={\i*\i/16}] (2*\L+2*\r-\x+1.4, 1.5*\h) circle (0.2);
	}
	\draw[->, thick] (\L,0) -++ (\r,0);
	\fill[Red] (2*\L+2*\r+1.9, 1.5*\h) circle (0.2);
	
	\draw (2*\L+2*\r,-1.5*\h-\h) rectangle ++(\L,\h*2) node[midway,below,inner sep=\sep cm] {IIIb};
	\draw[->,thick] (3*\L+2*\r,-1.5*\h) -++ (\r,1.5*\h);
	\fill[Blue] (2*\L+2*\r+1.5, -1.5*\h) circle (0.2);
		\foreach[count=\i] \x in {0, 0.1, 0.2, 0.3}
	{
		\fill[Red, opacity={\i*\i/16}] (2*\L+2*\r+\x+1.9, -1.5*\h) circle (0.2);
	}
	
	\draw (3*\L+3*\r,-\h) rectangle ++(\L,\h*2) node[midway,above,inner sep=\sep cm] {IV};
	\foreach[count=\i] \x in {0, 0.1, 0.2, 0.3}
	{
		\fill[Blue, opacity={\i*\i/16}] (3*\L+3*\r+\x+1.7, 0) circle (0.2);
	}
	\foreach[count=\i] \x in {0, 0.1, 0.2}
	{
		\fill[Red, opacity={1-(\i-2)*(\i-2)/2}] (3*\L+3*\r+2.3+\x, 0) circle (0.2);
	}
	
	\end{tikzpicture}
	\caption{In this illustration of the inter-particle ratchet effect, the blue particle (`shepherd') has positive drift and small diffusivity; the red particle (`sheep') has zero drift and higher diffusivity. I$\to$II: the shepherd catches up with the sheep. II$\to$III: a fluctuation creates a separation, which is more likely the sheep meandering forward (IIIb) than the shepherd backward (IIIa). IIIb$\to$IV: shepherd again catches up. Through rectification, the sheep gets a non-zero net velocity.  }\label{fig:shepherd}
\end{figure}

We seek the density $\rho(x)$ of particles a clockwise distance from the dominant particle one. It is possible, but arduous, to perform the necessary marginalization over the joint position distribution . Fortunately, the exact result can be obtained by the following heuristic argument. We replace the dominant particle with a hard wall moving at fixed velocity $\tilde{v}$. A single $(v,D)$--particle trapped between two hard walls moving at velocity $\tilde{v}$ would have spatial distribution $ \propto \exp[ -(v - \tilde{v})/D)x] = \exp[-kx]$, where $x$ is the position relative to the left wall. By a certain reflection-symmetry of Brownian trajectories \cite{Rodenbeck1998}, identical particles in single-file behave as a set of non-interacting particles. We therefore expect the density 
\begin{equation}\label{eq:rho(x)}
\rho(x) = (N-1) \frac{k e^{-kx}}{1- e^{kL}},
\end{equation}
which can be proved exactly from \eqref{eq:SFD-P-y} (cf.\ the main calculations in \autoref{sec:tracers}) . Without loss of generality we assume $v_1 > v$ so that $1/k >0$ gives the characteristic length scale over which particles are clustered  ahead of particle one. It is thus apparent that when diffusion enters the picture, a `shepherd' particle that is both fast and has low diffusivity can force `sheep' particles ahead of it to speed up through a ratchet effect: if a gap opens up between the shepherd and the sheep, it is most likely to be because a sheep diffuses forward than the shepherd diffusing backward. The shepherd is then quick to close the gap. This is illustrated in \autoref{fig:shepherd}. Similarly, a stubborn `donkey' particle with  speed $v_1 < v$ and low diffusion will slow down the faster `horse carriages' behind it if the horses are more diffusive.

As a final comment, if we consider the quasi- as opposed to exactly one-dimensional single-file diffusion, with isotropic diffusivity $D_i$, then with recourse to \eqref{eq:intro-nu}
\begin{equation}
\tilde{v} = \sum_i \alpha_i \hat{\vec{r}}\cdot \vec{v}^{(i)},
\end{equation}
with $\alpha_i$ as defined above. The conclusions about the effects of disorder will thus be quantitatively the same in the quasi-one-dimensional case.

\subsection{Driven tracers in a passive bath}\label{sec:tracers}

A typical scenario in statistical mechanics is measuring the response of a system when a force is applied to a selected particle---a tracer---in a collection of otherwise identical particles. For an equilibrium system experiencing a small perturbation, classical linear response theory governs the outcome. Out of equilibrium, an array of intriguing response phenomena are possible. For example, in the case of a single driven tracer which can overtake with some rate, this parameter separates phases in the tracer current \cite{Miron2020}, and can enable absolute negative mobility \cite{Civindi2018}.

A recent paper \cite{Lobaskin2020} studied the steady state properties of one or more driven tracers in a bath of passive particles, in the absence of any overtaking. The tracers are totally asymmetric random walkers on a periodic lattice, and the bath particles are symmetric. For a single driven tracer, there is an exponentially decaying pile-up of bath particles ahead of it (cf.\ \eqref{eq:rho(x)} with $v = 0$, $D > 0$ for the bath particles and $v_1 > 0$, $D_1 = 0$ for the tracer). For a system with several tracers, a cooperation effect was described where the system current grows with the density of tracers.

We can view the dichotomy of tracer/non-tracer particles as an instance of particle heterogeneity, and utilize the results of the previous section to analyse the continuum-version of driven tracers in a passive bath.  Strictly speaking, the continuum-limit of the lattice model \cite{Lobaskin2020} would have the tracers be deterministic, and since the bath particles diffuse, ratcheting would force the current to be exactly the tracer speed $v_T$. Instead, let us suppose both tracers and bath particles have identical diffusivity $D$. Then if $\rho_T = N_T / N$ is the fraction of all particles that are tracers, the common net velocity of the system will according to \eqref{eq:SFD-nu-weighted} be
 \begin{equation}\label{eq:nu=rho_tv_T}
 \tilde{v} = \rho_T v_T,
 \end{equation}
 since the bath particles have zero intrinsic velocity. Here, it does not matter for the current where amongst the bath particles the driven tracers are placed: because all particles are diffusive, tracers can ratchet each other if they are adjacent, just as well as they can ratchet bath particles. 

In the remainder of this section we derive an exact expression for  the joint   distribution  of gaps between the  driven tracers by integrating out the bath particles in the steady state density \eqref{eq:SFD-P-y} for arbitrary disorder. As we shall see, the exact solution for the driven tracers---which we emphasize represents the steady state of a system driven arbitrarily far from equilibrium---can be interpreted as a subsystem of interacting \textit{heterogeneous} tracers, and subsystems of bath particles self-distributing in boxes (the space between the tracers), with these subsystems conjoined through the conservation of space and Eq.\ \eqref{eq:nu=rho_tv_T}. The effective heterogeneity between tracers arises when they differ in the number of bath particles ahead of each one before the next tracer. 

In preparation for this calculation, let us write down the marginalized density for a selected subset of particles, even for full particle-wise disorder. Select a subset of $M$ particles, with index $\tau_1, \ldots, \tau_M$, and let $z_i$ be the gap between particle $\tau_i$ and $\tau_{i+1}$. Using the colon summation notation \eqref{eq:sumnot},
\begin{equation}
z_i = y_{\tau_i : \tau_{i+1}-1},
\end{equation}
with $y_i$ the gap between particle $i$ and $i+1$, as before. Then
\begin{subequations}\label{eq:Ptau}
	\begin{align}
	P_{\{\tau\}}&(z_1,\ldots,z_M)  = \avg{\prod_{i=1}^M \delta\left( y_{\tau_i : \tau_{i+1}-1} - z_i  \right)} \\
	&= \frac{1}{Z_N(w_1,\ldots,w_N;L)} \left(\prod_{i=1}^N \int_0^L dy_i\, e^{ w_i y_i}\right)  \delta\left( L - y_{1:N} \right) \prod_{i=1}^M \delta\left( y_{\tau_i : \tau_{i+1}-1} - z_i  \right) \\
	& =\delta\left( L - z_{1:M} \right) \frac{\prod_{i = 1}^M Z_{\tau_{i+1}- \tau_i} (w_{\tau_i}, \ldots, w_{\tau_{i+1}-1}; z_i) }{Z_N(w_1,\ldots,w_N;L)}.
	\end{align}
\end{subequations}
To get to the last line, we note that the product of delta functions decouple the integrals into ``blocks'' $ [ \tau_{1}, \tau_1 + 1  \ldots \tau_2 -1 ] $ $[\tau_{2}, \ldots ] \cdots [\tau_M, \ldots, N]$, each of which by definition recovers a partition function \eqref{eq:Z-def} for the appropriate arguments.
\begin{figure}
	\centering
	\begin{tikzpicture}
	
	\def \tcol {Blue!60!White}
	\def \bcol {Red!60!White}
	\def \zone {2.8}
	\def \ztwo {1.6}
	\def \gap {1.5}
	\def \zN {2.2}
	
	\draw[|-|] (0,-0.2)--(\zone,-0.2) node[midway,below] {$z_1$};
	\draw[fill,\tcol] (0,0.2) circle (0.2);
	\node at (\zone/2-0.17,0.8) {$\overbrace{\hspace{1.95 cm}}^{b_1=4}$};
	\foreach \x in {0.15,0.35,0.51,0.72}
	{
		\draw[fill,\bcol] ({\x*\zone} ,0.2) circle (0.2);
	}
	
	\draw[-|] (\zone,-0.2)--({\zone+\ztwo},-0.2) node[midway,below] {$z_2$};
	\draw[fill,\tcol] (\zone,0.2) circle (0.2);
	\draw[fill,White] (0,-0.15) rectangle ++(-0.3,0.6);
	\node at ({\zone+\ztwo/2-0.15},0.8) {$\overbrace{\hspace{0.85cm}}^{b_2=2}$};
	\foreach \x in {0.29,0.57}
	{
		\draw[fill,\bcol] ({\zone+\x*\ztwo} ,0.2) circle (0.2);
	}
	\draw[fill,\tcol] ({\zone+\ztwo},0.2) circle (0.2);

	\draw[dashed] ({\zone+\ztwo},-0.2) --+ (\gap,0);
	\node at ({\zone+\ztwo+\gap/2},0.2) {$\cdots$};
	
	\draw[|-] ({\zone+\ztwo+\gap},-0.2)--+ (\zN,0) node[midway,below] {$z_{N_T}$};
	\draw[fill,\tcol] ({\zone+\ztwo+\gap},0.2) circle (0.2);
	\node at ({\zone+\ztwo+\gap+\zN/2+0.07},0.8) {$\overbrace{\hspace{1.6 cm}}^{b_{N_T} = 3}$};
	\foreach \x in {0.24,0.44,0.81}
	{
		\draw[fill,\bcol] ({\zone+\ztwo+\gap+\x*\zN} ,0.2) circle (0.2);
	}
	\draw[fill,\tcol] ({\zone+\ztwo+\gap+\zN},0.2) circle (0.2);
	\draw[fill,White] ({\zone+\ztwo+\gap+\zN},-0.15) rectangle ++(0.3,0.6);	

	\end{tikzpicture}
	\caption{Sketch of driven tracers in a passive medium. Blue balls are tracers and red balls are passive bath particles. The  distance between a tracer $i$ and the next is $z_i$, and there is a number $b_i$ of bath particles in between. }\label{fig:TBBB}
\end{figure}
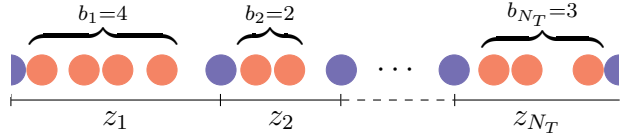

Now, let the particle $\tau_n$ be the $n$th tracer, in front of which there are $b_n$ bath particles until the next tracer, as shown in \autoref{fig:TBBB}. The total number of tracers is $N_T$ and the bath particles number $N_B = b_{1:N_T} = N - N_T$. Recall the definitions of $w_i = k_{i+1:N}$ \eqref{eq:w_i} and $k_i = (v_i - \tilde{v})/D_i$ \eqref{eq:k_i}. For a two-species setup, $k_i$ is either 
	\begin{equation}\label{eq:kBkT}
	k_B \equiv -k\quad \text{or} \quad  k_T = (N_B / N_T ) k.
	\end{equation}
Since $k$ is the only continuous parameter on which the distribution depends, we can without loss of generality assume both particles types to have the same diffusivity $D$, whereas $v_T > 0$, and $v_B = 0$. Then
\begin{equation}
k = \tilde{v} / D = \rho_T v_T / D  > 0. 
\end{equation}
We need to calculate partition functions of the form
\begin{equation}\label{eq:Z_bm}
Z_{b_n + 1}(w_{\tau_n}, \ldots, w_{\tau_n + b_n}; z_n) = \sum_{i = 0}^{b_n} \exp[{w_{\tau_n + i}\, z_n}]  \prod_{j = 0 \neq i}^{b_n} \frac {1}{w_{\tau_n + i} - w_{\tau_n + j}}.
\end{equation}
We write the weight factor in the exponential as
\begin{alignat}{2}\nonumber
w_{\tau_n + i} = k_{\tau_n +i + 1 : N} &= k_{\tau_n + i+1 : \tau_{n+1}-1} && + k_{\tau_{n+1}:N} \\
& = (-k) (b_n - i) && + (- k) g_n. 
\end{alignat}
By definition, $-kg_n = (N_T - n) k_T + b_{n+1:N_T} k_B$, which using the expressions for $k_T$ and $k_B$ \eqref{eq:kBkT} can be written more meaningfully as 
\begin{equation}
g_n = \bar{b} n - b_{1:n} = \sum_{m = 1}^n (\bar{b} - b_{m}),
\end{equation}
where $\bar{b} = N_B / N_T$ is the average number of consecutive bath particles. Next, the difference of weight factors in the product in \eqref{eq:Z_bm} can be written
\begin{equation}
w_{\tau_n + i} - w_{\tau_n + j} = (-k) (j - i).
\end{equation}
Hence
\begin{equation}
\prod_{j = 0 \neq i}^{b_n} \frac {1}{w_{\tau_n + i} - w_{\tau_n + j}} =  \frac{1}{k^{b_n}} \frac{(-1)^{b_n - i}}{i! (b_n - i)!}.
\end{equation}
Putting these results together,
\begin{subequations}
\begin{align}
Z_{b_n + 1} &= \frac{e^{-k g_n z_n}}{k^{b_n} b_n!} \sum_{i=0}^{b_n} (-1)^{b_n - i} e^{- k(b_n - i) z_n} {{b_n}\choose{i}}  \\
&= \frac{ e^{-k g_n z_n} }{b_n!} \left( \frac{1 - e^{-k z_n}}{k} \right)^{b_n} .
\end{align}
\end{subequations}
Finally, the tracer gap distribution comes out as 
\begin{equation}\label{eq:Pz}
P_{\{\tau\}}(z_1,\ldots,z_{N_T}) = \delta\left( L - \sum_{n=1}^{N_T} z_{n} \right)  \frac{1}{Z_N} \prod_{n=1}^{N_T}\frac{ e^{-k g_n z_n} }{b_n!} \left(  \frac{1- e^{-k z_n} }{k} \right)^{b_n}.
\end{equation}
This is the main result of this section. We now turn to its interpretation.

We first note that expression \eqref{eq:Pz} factorises over the tracer particles labelled by $n$.
Each factor  involves only the gap size $z_n$ to the next tracer.
Note, however, that the gap sizes are not independent due to the global constraint that their sum 
must be equal to the system length $L$, which is implied by the delta function.

Let us refer to each factor in the product as a gap size weight. Each contains an exponential factor  $\exp[ - k  g_n z_n]$ and a factor involving $b_n$ (the number of bath particles between tracer $n$ and $n+1$), which we now consider in turn. By themselves, the exponential factors give $\exp[ - k \sum_n g_n z_n]$  which has (with gap conservation and appropriate normalization)  the form of the gap distribution \eqref{eq:SFD-P-y} pre-marginalization. That hypothetical gap distribution would formally describe a system of $N_T$ interacting {\em heterogeneous} tracers, each with its own effective velocity, with no surrounding bath. Comparing the exponential constant $-k g_n$ with $w_n$ in \eqref{eq:SFD-P-y} for fixed diffusivity $D$  one deduces that the effective velocity of tracer $n$ would be 
\begin{equation}
u_n = u^\text{ref} -\tilde{v} b_n =  u^\text{ref} - [b_n/(1 + \bar{b})] v_T  ,
\end{equation}
where $u^\text{ref}$ is some arbitrary reference velocity. A tracer with fewer bath particles ahead of it with respect to the actual model than the average tracer, would here have larger (signed) effective velocity $u_n$. As a result, it will disproportionally compress the space to its neighbours in the positive direction. This effect grows with $k$. For instance, taking $N_T = 2$, if $b_1 > b_2$, then typically $z_1$ will be large, as tracer two with fewer bath particles in front of it succeeds in compressing the relatively fewer bath particles in front of it until it gets close to tracer one. 

Note that if all $b_n = N_B/N_T$  there is no heterogeneity and  all $g_n = 0$. As one then expects from symmetry, typically all $z_n \approx L/N_T$ and the remaining factor involving $b_n$ in \eqref{eq:Pz} determines the distribution.

To interpret this remaining factor, consider a single bath particle trapped in a box $[0, L]$ whose boundaries move forward at velocity $\tilde{v}$. Its position with respect to the left box boundary would be distributed with cumulative density $q_B(x) \propto  1 - \exp[-kx]$. The probability of finding $b_n$ indistinguishable particles crammed into $[0,z_n]$ would be
\begin{equation}
q_B^{b_n}(z_n) / b_n!\ ,
\end{equation}
which is precisely the factor appearing in the gap size weight.

In conclusion, the full gap distribution \eqref{eq:Pz} may be
viewed as a product of two subsystems (each of which factorises):  one describing a system of tracers without bath particles with heterogeneous effective velocities, and one representing a system of  bath particles distributed in boxes (the space between a pair of tracers). 
The weight for a single gap size $z_n$ is then the product of the weight for  gap size $z_n$ in the tracer subsystem multiplied by the probability that $b_n$ bath particles occupy space less than 
$z_n$ in the bath subsystem. 
However, these subsystems are correlated through the conservation of space and the global current established cooperatively by all particles. 
 
\subsection{The Brownian many-filament ratchet}\label{sec:ratchet}

\begin{figure}
	\centering
		\begin{tikzpicture}
		\def \R {1cm}
		\def \n {10}
		\foreach \p in {1,...,\n} 
		{
			\draw[dashed] ({\R*cos(360/\n*(\p+0.5))}, {\R*sin(360/\n*(\p+0.5))}) circle ({pi*\R/\n});
		}
		\draw[color=black,fill=white] (0,0) circle (\R);
		\foreach \p in {1,...,\n} 
		{
			\draw[fill=black] ({\R*cos(360/\n*\p)}, {\R*sin(360/\n*\p)}) circle (0.05cm);
		}
		\node at (0,-3) {(a) Periodic SFD};
		\end{tikzpicture}
		\hspace{2cm}
		\begin{tikzpicture}
		\def \n {5}
		\def \L {3cm}
		\def \xlist {3.1,1.5,2.4,2.9,0.8}
		\def \xpiston {3.3}
		\def \xmin {0}
		\def \xmax {3.8}
		\def \yoff {0.4};
		
		\draw (\xmin,0) -- (\xmax,0); 
		
		\foreach[count=\i] \x in \xlist {
			\draw[dotted] (\x,\yoff-0.2+\i*0.5)--(\x,0);
		}		
		
		\foreach[count=\i] \x in \xlist {
			
			\draw[fill=Blue!20!white] (\xmin,\yoff+\i*0.5)--(\x,\yoff+\i*0.5) --++(0,-0.2) --++({-\x},0);
			
			\draw[dashed] (\x,0) arc (-180:0:{(\xpiston-\x)/2});
			
			\draw[color=Blue,fill=Blue] (\x,0) circle (0.05);	
		}
	
		\fill[Red!20!White] (\xpiston,\yoff) rectangle ++(0.2,{(\n+1.2)*0.5-\yoff});
		
		\draw[dotted] (\xpiston,\yoff)--(\xpiston,0);
		\draw (\xpiston,{(\n+1.2)*0.5}) -- (\xpiston,\yoff);		
		\draw (\xpiston+0.2,{(\n+1.2)*0.5}) -- (\xpiston+0.2,\yoff);
		\draw[color=black,fill=Red] (\xpiston,0) circle (0.05cm);
		
		\node at (1,3.5) {filaments};
		\node at (3.4,3.5) {membrane};
		
		\node[] at ({(\xmax-\xmin)/2}, -2) {(b) Filament ratchet};		
		\end{tikzpicture}
		
	\caption{In SFD no particle can cross its neighbour. In the $N$-filament ratchet, we put the absolute length $x_i$ of the $N$ filaments as mutually non-interacting particles on the real line. Each filament however cannot cross through the membrane, represented as particle $N+1$. }\label{fig:SFD-ratchet}
\end{figure}
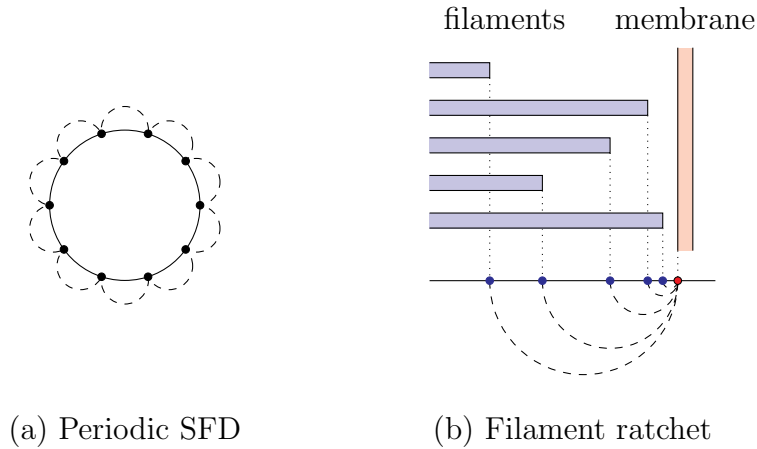

A recent paper by Wood et al.\ \cite{Wood2019} introduced a model for membrane growth, where a number of filaments grow by polymerization and extend a membrane in the growth direction through a ratchet effect (\autoref{fig:SFD-ratchet} (b)). A central result of that work was the solution for the net velocity $\tilde{v}_\text{M}$ of the membrane, when the $N$ filaments have constant nominal growth rates $v_{\text{F},i}$ with a diffusivity $D_{\text{F},i}$, whereas the membrane has a nominal velocity $-v_\text{M}$ and diffusivity $D_\text{M}$:   
\begin{equation}
\tilde{v}_{\text{M}} = \frac{ -v_{\text{M}} D_{\text{M}}^{-1} + \sum_{i=1}^N v_{\text{F},i}^{(i)} D_{\text{F},i}^{-1}  }{ D_{\text{M}}^{-1} + \sum_{i=1}^N  D_{\text{F},i}^{-1} }.
\end{equation}
This expression is formally identical to \eqref{eq:SFD-nu-1} derived for single-file diffusion. This ultimately follows from the fact that in both models, all degrees of freedom must have a common net velocity for geometric reasons.

\autoref{fig:SFD-ratchet} shows how to map the Brownian many-filament ratchet to a particle problem directly comparable to SFD. The absolute positions of the membrane and each filament correspond to the positions of particles on an infinite line. Two particles cannot pass one another if they are connected by a dashed line. For a periodic domain, it is intuitively clear that if the \emph{graph of mutual exclusions} (dots and dashed lines in \autoref{fig:SFD-ratchet}) includes all particles---whether any given two particles are directly connected---then they are geometrically constrained to have a common net velocity. This is true also for the infinite line, with the caveat that the model parameters must allow the particles to cluster rather than disperse in the long-time limit.

From a mathematical point of view, if we make the ansatz $J_i(\vx) = \tilde{v}_i P(\vx)$, then the no-crossing condition \eqref{eq:SFD-excl} for particles $i$ and $j$ immediately implies $\tilde{v}_i = \tilde{v}_j$, so that all particles in the graph of mutual exclusion must have the same $\tilde{v}$. If we define $\vu(\vx)$ by $\vec{J}(\vx) = \vu(\vx) P(\vx)$ for any steady state, then $\vu(\vx)$ is known as the irreversible drift (provided $\vx$ is even under time-reversal in all its components). In the version of SFD treated here, and in related cases, $\vu$ is a constant vector. The significance of this is that one can change variables into a frame moving through phase space with velocity $\vu$, relative to which the process satisfies detailed balance---this explains the solvability of the model. This is more than an ``equilibrium system on wheels'', however, as the velocity of the metaphorical cart arises non-trivially from microscopic interactions. We expound the role of the irreversible drift, and its relation to phase space structure, in \autoref{sec:defsol}.

\section{$N$-body volume exclusion process in a $d$-dimensional tube: integrability conditions and steady-state solution}\label{sec:defsol}

\subsection{Phase space structure and stochastic dynamics}\label{sec:defs}

\begin{figure}
	\centering
	\begin{tikzpicture}
	\draw (0,0)--(4,0);
	\draw (2,-0.5) circle (0.5);
	\draw[dashed] (2,-1.5)--(2,0.5);
	\node at (3.15,-0.4) {$r_i \delta \hat{\vq}$};
	\node at (0.6,-0.4) {$\vq$};
	\node at (0.7,-1.1) {$\vq^{(i)}$};
	\draw[-{Latex[length=3mm, width=1.5mm]}] (0,-1)--(2,-0.5) ;
	\draw[-{Latex[length=3mm, width=1.5mm]}] (0,-1)--(2,0) ;
	\draw[-{Latex[length=3mm, width=1.5mm]}] (2,0)--(2,-0.5) ;
	\foreach \x in {0,0.2,...,3.8} {
		\draw (\x,0)--(\x+0.15,0.15);
	}
	\draw[gray] (2.1,-0.35) -- (2.65,-0.4);
	\draw[-{Latex[length=3mm, width=1.5mm]}] (1.5,-2)--(2.5,-2) node[anchor=west] {$\hat{\vec{r}}$} ;
	\end{tikzpicture}\qquad%
	\begin{tikzpicture}
	\draw[dashed] (0,0) -- (2,-3);
	\draw (2/3,-1) circle (0.5);
	\draw (4/3,-2) circle ({sqrt(13)/3-0.5});
	\draw[-{Latex[length=3mm, width=1.5mm]}] (-0.5,-2.5)--(4/3,-2) ;
	\draw[-{Latex[length=3mm, width=1.5mm]}] (-0.5,-2.5)--(2/3,-1) ;
	\draw[-{Latex[length=3mm, width=1.5mm]}] (4/3,-2)--(2/3,-1) ;
	\node at (2.4,-0.9) {$(r_i+r_j) \delta \hat{\vq}$};
	\draw[gray] (1.2,-1.65)--(2.25,-1.15);
	\node at (-0.3,-1.7) {$\vq^{(i)}$};
	\node at (0.4,-2.6) {$\vq^{(j)}$};		
	\end{tikzpicture}
	\caption{Direction of incidence $\delta \hat{\vq}$ for collision between particle $i$ and a wall or between particle $i$ and particle $j$.}\label{fig:coll}
\end{figure}
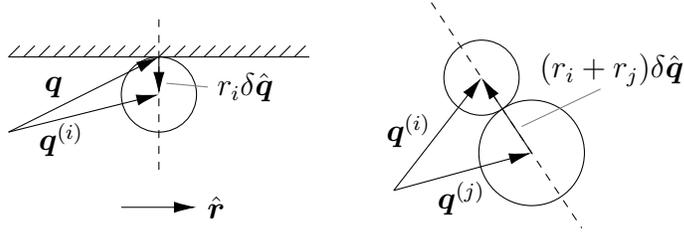

The physical meaning and broad mathematical structure of the general model were described in the \autoref{sec:model}. Here we give the further details. The drift-diffusion process $\mathbf{X}(t)$ evolves in a phase space  $\Gamma \subset \mathbb{R}^{d \times N}$ constructed as follows. The tube is a space $\mathcal{Q} \subset \mathbb{R}^d$ within which there is a unique `axial' direction $\hat{\vec{r}}$ that is unbounded whereas all other directions are bounded. The boundary $\del \mathcal{Q}$ is axially constant, meaning that a boundary normal $\hat{\vec{n}}$ is always orthogonal to $\hat{\vec{r}}$. Each particle $i \in \{1,2,\ldots,N \}$ inherits a copy $\mathcal{Q}^{(i)}$ of $\mathcal{Q}$ so $\Gamma \subset \mathcal{Q}^{(1)} \times \cdots \times \mathcal{Q}^{(N)}$. The inclusion is strict because we have yet to exclude phase space points forbidden due to physical volume exclusion. Recall that the components of $\vx \in \Gamma$ are organized according to 
\eqref{eq:X} as
\begin{equation}
\vx = \sum_{i=1}^N \hat{\vec{e}}_i \otimes
\vq^{(i)},
\end{equation}
with $\vq^{(i)} \in \mathcal{Q}^{(i)}$. The part of the phase space boundary $\del \Gamma$ arising from particle $i$ touching the tube walls at $\vq \in \del \mathcal{Q}$ is  
\begin{equation}
B_i(\vq) = \{ \vx \in \Gamma: ||\vq^{(i)} - \vq|| = r_i \},
\end{equation}
where $r_i$ is the radius specific to particle $i$. For future reference we compute the boundary normal by applying the gradient operator 
\begin{equation}
\nabla \equiv \nabla_{\vx} = \sum_{i=1}^N \hat{\vec{e}}_i \otimes 
\nabla_{\vq^{(i)}}
\end{equation}
to the  locus $||\vq^{(i)} - \vq|| = r_i$. The result is
\begin{equation}\label{eq:ni}
\hat{\vec{n}}(\vx) = \hat{\vec{e}}_i \otimes 
\delta \hat{\vq}(\vx),\quad \vx \in B_i(\vq),
\end{equation} 
with $\delta \hat{\vq} = (\vq^{(i)} - \vq)/r_i $ as illustrated in \autoref{fig:coll} (left panel). Note that $\delta \hat{\vq} \cdot \hat{\vec r} = 0$.

For mutual volume exclusion between particles we consider an `interaction graph' $G$ where a node represents a particle, and an edge between two particles signifies that they mutually exclude volume. E.g., SFD has the complete graph, and the filament ratchet a star graph (see \autoref{fig:SFD-ratchet}). If $(i,j) \in G$ then we must introduce a boundary surface
\begin{equation}
B_{ij} = \{ \vx \in \Gamma: ||\vq^{(i)}-\vq^{(j)}|| = r_i + r_j \}.
\end{equation}
Its normal is
\begin{equation}\label{eq:nij}
\hat{\vec{n}}(\vx) = \frac{1}{\sqrt{2}} (\hat{\vec{e}}_i - \hat{\vec{e}}_j) \otimes \delta \hat{\vq}(\vx),\quad \vx \in B_{ij},
\end{equation}
where $\delta \hat{\vq} \propto \vq^{(i)}-\vq^{(j)}$, illustrated in \autoref{fig:coll} (right panel). 

The full phase space boundary is then
\begin{equation}
\del \Gamma = \left(\bigcup_{i,\vq \in \del Q} B_i(\vq) \right)\bigcup \left(\bigcup_{(i,j) \in G} B_{ij} \right).
\end{equation}
$\Gamma$ is the restriction of $\mathbb{R}^{d\times N}$ that lies inside or on this boundary. If the geometry of the tube and particles is such that a certain ordering of particles is logically preserved, e.g.\ as is necessarily the case for SFD ($d=1$ and $G$ the complete graph) then the process is confined to the sector $\Gamma_{\vx_0} \subset \Gamma $ containing the initial condition $\vx_0$.

We now generalize the dynamics given in \eqref{eq:sde} to allow non-constant drift and diffusion parameters. In terms of a formal It\^o SDE,
\begin{equation}
d \mathbf{X}(t) = \vec{a}(\mathbf{X}(t)) dt + \mathsf{B}(\mathbf{X}(t)) d\mathbf{W}(t) + d \mathbf{R}(t),
\end{equation}
where $\mathbf{W}(t)$ is a $Nd$-dimensional Wiener process, and the exact definition of the reflection term $\mathbf{R}(t)$ will be provided in the Fokker-Planck picture. The  density $P(\vx,t)$ defined similar to \eqref{eq:P(x)}, given $\vx$ in the interior $\Gamma^{\text{int}} = \Gamma \setminus \del \Gamma$ of the phase space, evolves by the Fokker-Planck equation 
\begin{equation}
\del_t P(\vx,t) = - \sum_i \del_{x_i} a_i(\vx)P(\vx,t) + \frac{1}{2} \sum_{i,j} \del_{x_i} \del_{x_j}[\mathsf{B}(\vx) \mathsf{B}^\top(\vx)]_{ij} P(\vx,t).
\end{equation}
We find the following parameters more convenient:
\begin{equation}
v_i(\vx) =  a_i(\vx) - \frac{1}{2} \sum_j \del_{x_j} [\mathsf{B}(\vx) \mathsf{B}^\top(\vx)]_{ij},\quad \mathsf{D}(\vx) = \frac{1}{2} \mathsf{B}(\vx) \mathsf{B}^\top(\vx).
\end{equation}
We will refer to $\vv(\vx)$, rather than $\vec{a}(\vx)$, as the \emph{drift vector}, pointing out that this may break from common convention if the diffusion matrix $\mathsf{D}(\vx)$ is non-constant. For the latter we make the mild assumption that it is invertible. Since the diffusion matrix is a covariance matrix, it is symmetric and positive \emph{semi}-definite. Its invertibility is therefore equivalent to \emph{strict} positive definiteness. What this means is that there is no linear combination of the components of $\mathbf{X}(t)$ that has zero variance, hence evolves deterministically.

We then compactly write the FPE as
\begin{equation}\label{eq:FPE}
\del_t P(\vec x, t)  + \nabla \cdot \vec{J}(\vec{x},t) = 0,
\end{equation}
using the probability current
\begin{equation}\label{eq:Jt}
\vec{J}(\vec x,t) = \Vec{v}(\vec{x}) P(\vec{x},t) - \mathsf{D}(\vec x) \nabla P(\vec x,t). 
\end{equation}
On the phase space boundary $\del\Gamma$ the process is reflected. This means that the probability flow through boundaries must vanish:
\begin{equation}\label{eq:J.n}
\vec{J}(\vec{x},t) \cdot \hat{\vec{n}}(\vec{x}) = 0,\quad \vec{x} \in \del \Gamma.
\end{equation}

The remaining conditions differ depending on whether the tube is periodic or infinite. In the following we will suppose periodicity, and leave the case of infinite tube to \autoref{sec:inf}. Thus we require invariance under translating the whole system by one axial period $L$,
\begin{equation}\label{eq:globalperiod}
P(\vx + L \vtau,t) = P(\vx, t).
\end{equation}
Recall that $\vtau$ is the global translation vector
\begin{equation}
\vtau = \sum_{i=1}^N \hat{\vtau}^{(i)} = \vec{1} \otimes \hat{\vec r},
\end{equation}
constructed from the vectors
\begin{equation}\label{eq:taui}
\hat{\vtau}^{(i)} = \hat{\vec{e}}_i \otimes \hat{\vec r}
\end{equation}
that move particle $i$ forward axially a unit distance, keeping all else constant. By assuming more strongly translation invariance of the model parameters, i.e.\ axial homogeneity of the tube,
\begin{equation}
\vv(\vx + r {\vtau}) =  \vv(\vx),\quad \mathsf{D}(\vx + r {\vtau}) = \mathsf{D}(\vx),\quad r \in \mathbb{R},
\end{equation}
we obtain \eqref{eq:globalperiod} automatically.

As mentioned in the introduction, further conditions appear if the geometry is not quasi-one-dimensional. Suppose that there exists some \emph{possible} trajectory $\omega = \{\vec{x}(s) \}_{s=0}^{T}$, where $\vx(T) = \vec{x}(0) + L\hat{\vtau}^{(i)}$. That is, particle $i$ has made one net circulation of the tube, while the other particles have not. Then $\vx(0)$ and $\vx(T)$ are physically the same due to tube periodicity and therefore we must require,
\begin{equation}\label{eq:localperiod}
P(\vx + L\hat{\vtau}^{(i)},t) = P(\vx, t).
\end{equation}
However, if no such trajectory exists for particle $i$, i.e.\ it is prevented by other particles from circling the tube by itself, then there is no logical reason to impose \eqref{eq:localperiod}.

\subsection{Conditions for a solvable steady state}\label{sec:sol}

It is known in the literature that there exists certain ``potential conditions'' \cite{Graham1971,SanMiguel1980,Cai1988} which make the steady state of a drift-diffusion solvable in terms of an explicit integral---this is what we mean by integrability in this article. In this section we derive these conditions in an unconventional way, tailored to the presence of a phase space boundary. A steady state density $P(\vx)$ for the process $\mathbf{X}(t)$ in its original coordinates exists, as guaranteed by the boundedness of the phase space of the periodic tube. Let us define the function $\vec{u}(\vec{x})$ by
\begin{equation}\label{eq:J=uP}
\vec{J}(\vec{x}) = \vec{u}(\vec{x}) P(\vec{x}).
\end{equation}
Assuming $P(\vx)>0$ for all $\vx$ this definition is unambiguous. We take the components of $\mathbf{X}(t)$ to be even under time reversal. Then $\vu(\vx)$ is called the irreversible drift \cite{Graham1971} for reasons to become clear. Eliminating $\vec{J}$ for $\vu$ in the stationary versions of \eqref{eq:FPE}, \eqref{eq:Jt}, and \eqref{eq:J.n} the result is
\begin{subnumcases}{\label{eq:usys}}
\nabla \cdot \vec{u}(\vec{x}) + \vec{u}(\vec{x})\cdot \nabla \ln P(\vec{x}) = 0, & $\vec{x} \in \Gamma$, \label{eq:divu}\\
\nabla \ln P(\vec{x}) = \mathsf{D}^{-1}(\vec x) [ \vec{v}(\vec{x}) - \vec{u}(\vec{x})], & $\vec{x} \in \Gamma$,\label{eq:dlnP}\\
\vec{u}(\vec{x})\cdot \hat{\vec n}(\vec{x}) = 0, & $\vec{x} \in \del \Gamma$. \label{eq:u.n}
\end{subnumcases}
By combining \eqref{eq:divu} and \eqref{eq:dlnP} we can view \eqref{eq:usys} as a closed equation for $\vu(\vx)$, 
\begin{subnumcases}{\label{eq:vusys}}
\nabla \cdot \vec{u}(\vec{x}) + \vec{u}(\vec{x})\cdot \mathsf{D}^{-1}(\vec x) [ \vec{v}(\vec{x}) - \vec{u}(\vec{x})], & $\vec{x} \in \Gamma$, \label{eq:vueq}\\
\vec{u}(\vec{x})\cdot \hat{\vec n}(\vec{x}) = 0, & $\vec{x} \in \del \Gamma$, \label{eq:u.n2}
\end{subnumcases}
together with the definition of a potential $\Phi_{\vec{ u}}(\vec{x})$ by
\begin{equation}\label{eq:Phi_u}
\nabla \Phi_{\vec{u}}(\vec{x}) = \mathsf{D}^{-1}(\vec x) [ \vec{v}(\vec{x}) - \vec{u}(\vec{x})].
\end{equation}
This potential generates the probability density through
\begin{equation}\label{eq:P}
P(\vx) = \frac{e^{\Phi_{\vu}(\vx)}}{Z},
\end{equation}
following the path-independent integration
\begin{equation}\label{eq:potint}
\Phi_{\vu}(\vx) - \Phi_{\vu}(\vec{c}) = \int_{\vec{c}}^{\vx} d\vec{y}\cdot \nabla \Phi_{\vec{u}}(\vec{y}).
\end{equation}
The assumed existence and uniqueness of the steady state has two important consequences: For given $\vv(\vx)$ and $\mathsf{D}(\vx)$, the problem \eqref{eq:vusys} (together with any additional boundary conditions, like periodicity) has a unique solution $\vu(\vx)$, and this solution necessarily makes \eqref{eq:Phi_u} integrable. 

But we can turn the logic around. Instead of fixing  $\vv(\vx)$ and $\mathsf{D}(\vx)$, which implies a specific $\vu(\vx)$, we assume that $\vu(\vx)$ is of a certain form, which imposes restrictions on which $\vv(\vx)$ and $\mathsf{D}(\vx)$ are consistent with this form. The constraints on $\vv(\vx)$ and $\mathsf{D}(\vx)$ are that they must solve \eqref{eq:vusys} for the given $\vu(\vx)$ while also making \eqref{eq:Phi_u} integrable. For example, $\vu = \vec{0}$ is always a trivial solution of \eqref{eq:vusys}. Equation \eqref{eq:Phi_u} then states the well-known integrability conditions for a detailed-balanced system. 

Observe that given the process $\mathbf{X}(t)$ with drift vector $\vv(\vx)$ and diffusion $\mathsf{D}(\vx)$ leading to irreversible drift $\vu(\vx)$, we can always define a new process $\mathbf{X}'(t)$ with the same steady state density but satisfying detailed balance: take  this process to have $\vv'(\vx) = \vv(\vx) - \vu(\vx)$ and $\mathsf{D}'(\vec{x}) = \mathsf{D}(\vec{x})$. The trivial solution $\vu' = \vec{0}$ is then most general one because it assumes no new integrability conditions on $\vv'(\vx)$ and $\mathsf{D}'(\vx)$ in order to solve for $\Phi'_{\vec{u}'}(\vx) $ ( $ = \Phi_{\vec{u}}(\vx) $). This shows that the irreversible drift $\vu(\vx)$ is the part of the drift without which detailed balance holds, but the density remains the same, hence its name.

To make progress with the tube problem, consider the following ansatz for the solution $\vu(\vx)$:
\begin{enumerate}[(i)]
	\item The direction $\hat{\vu}$ is constant
	\item The magnitude $u(\vx) \not\equiv 0$ is invariant in the $\hat{\vu}$-direction,
	\begin{equation}
	u(\vx) = u(\vx + s \hat{\vu})\quad \text{for all }s \in \mathbb{R}. 
	\end{equation}
\end{enumerate}
The condition (i) is motivated by the observation that there may exist such a \emph{constant} direction that solves all the reflective boundary conditions. We make the ansatz
\begin{equation}
\hat{\vu} \propto \sum_i \hat{u}_i \hat{\vtau}^{(i)}
\end{equation}
(recall \eqref{eq:taui}). Then $\hat{\vu}\cdot \hat{\vec n} = 0$ for the normal \eqref{eq:ni} of particle-wall boundary, and holds as well for the normal \eqref{eq:nij} of the particle$_i$-particle$_j$ boundary if
\begin{equation}
\hat{u}_i = \hat{u}_j.
\end{equation}
Let us assume that all particles are connected in the graph of mutual exclusions, but not necessarily that the graph is complete. Then it follows that $\hat{\vu} = \hat{\vtau}$.

The condition (ii) has the effect that $\nabla \cdot \vu(\vx) = \hat{\vu} \cdot \nabla u(\vx) = 0$. Then $u(\vx)$ can be solved for algebraically from \eqref{eq:vueq} as
\begin{equation}\label{eq:u(x)}
u(\vx) = \frac{\hat{\vu}^\top \mathsf{D}^{-1}(\vec{x}) \vec{v}(\vec{ x})}{\hat{\vu}^\top \mathsf{D}^{-1}(\vec{ x})\hat{\vu}  }.
\end{equation}
For condition (ii) to hold, we would generically require 
\begin{equation}\label{eq:vDtrans}
\vv(\vx + s \hat{\vu}) =  \vv(\vx),\quad \mathsf{D}(\vx + s \hat{\vu}) = \mathsf{D}(\vx),\quad \text{for all } s \in \mathbb{R}.
\end{equation}
Since  $\hat{\vu} = \hat{\vtau}$, the above simply states translation invariance which was already assumed.

With this trial solution, we have
\begin{equation}\label{eq:PhiH}
\nabla \Phi_{\vec{u}}(\vec{x}) = \mathsf{H}(\vx) \vv(\vx),\quad \mathsf{H}(\vx) = \mathsf{D}^{-1}(\vec{x}) - \frac{\mathsf{D}^{-1}(\vx) \hat{\vu} \otimes \hat{\vu}^\top\mathsf{D}^{-1}(\vx)}{\hat{\vu}^\top \mathsf{D}^{-1}(\vec{ x})\hat{\vu}}.
\end{equation}
The matrix $\mathsf{H}(\vx)$ is symmetric and $\mathsf{H}(\vx) \hat{\vu} = \vec{0}$. We have implicitly assumed that this potential exists, which may be difficult or artificial except for $\vv, \mathsf{D}$ constant. 

If the system is \emph{not} quasi-one-dimensional, then we must also satisfy \eqref{eq:localperiod}. For instance, with constant model parameters, we will find $P(\vx) \propto \exp[\vv^\top \mathsf{H}\vx]$. The condition is then that
\begin{equation}\label{eq:vHt}
\vv^\top \mathsf{H} \vtau^{(i)} = 0
\end{equation}
for every particle $i$ that is able to overtake every other particle. Generically, these constraints cannot be accommodated because they overdetermine the components of the irreversible drift. Logically, assumptions (i) or (ii) on $\vu$ will be violated. Then, we cannot solve \eqref{eq:usys} by algebraic means. A trivial case where \eqref{eq:vHt} can in fact be solved, is when the particles have the same drift $\vv = v \hat{\vu}$ which is then only in the axial direction. Then we have a flat steady state. As seen directly from \eqref{eq:usys}, a flat density $P = 1 / \Omega$ with current $\vv(\vx)/\Omega$ occurs whenever $\vv(\vx)\cdot \hat{\vec{ n}}(\vx) = 0$ and $\nabla\cdot \vv(\vx) = 0$, i.e.\ when $\vu(\vx) = \vv(\vx)$.

\subsection{Derivation of the current}\label{sec:currentderivation}
We now seek the expression for the net velocity $\tilde{v}_i$ of particle $i$. Up to a choice of units, net velocity and current are equivalent and we will use the terms interchangeably by setting $L=1$. Consider the hypersurface $S_r^{(i)}$ in $\Gamma$ given by $\vq^{(i)} \cdot \hat{\vec r} = r$ and whose normal is $\hat{\vec n} = \nabla (\vq^{(i)} \cdot \hat{\vec r}) = \hat{\vtau}^{(i)}$. It corresponds to all configurations where the $i$th particle is a distance $r$ down the tube. The net velocity at this axial distance is given by integrating the probability flow across $S_r^{(i)}$, i.e.\ by integrating $\vec{J}(\vx) \cdot \hat{\vec{n}}$ over all $\vx \in S_r^{(i)}$:
\begin{subequations}
	\begin{align}
	\tilde{v}_i &=  L \int_{S^{(i)}_r} d\vx\,\hat{\vtau}^{(i)} \cdot \vec{J}(\vec{x}) \\	
	&  =  L \int_\Gamma d\vx\, \delta(\vq^{(i)} \cdot \hat{\vec r} - r) \hat{\vtau}^{(i)} \cdot \vec{J}(\vec{x}).
	\end{align}
\end{subequations}
Using the translation invariance of the system, 
\begin{subequations}
	\begin{align}
	\tilde{v}_i & = \frac{1}{L} \int_0^L dr\, \tilde{v}_i(r) \\
	& =  \int_\Gamma d\vx\,  \hat{\vtau}^{(i)} \cdot \vec{J}(\vec{x}) \\ & = \avg{\hat{\vtau}^{(i)} \cdot \vu(\vx)}.
	\end{align}
\end{subequations}
When, as we found before, $\vu = u \hat{\vtau}$, all $\tilde{v}_i = \tilde{v}$ are identical, with the final result 
\begin{equation}\label{eq:nuconst}
\tilde{v} =  \frac{\vtau^\top \mathsf{D}^{-1} \vec{v}}{\vtau^\top \mathsf{D}^{-1}\vtau  }.
\end{equation}

This case has a special interpretation because it represents the situation where a change of variables $\mathbf{X}(t) \to \mathbf{X}'(t) = \mathbf{X}(t) - \vu t$ into a moving frame brings about detailed balance relative to it.

\subsection{Extension to infinite tube}\label{sec:inf}
\begin{figure}
	\centering
	\newcommand{\particle}[4]{
		\draw[->] #1 -- +({(#2+#3) * cos(#4)},{(#2+#3) * sin(#4)});
		\draw[fill,white] #1 circle (#2);
		\draw #1 circle (#2);
	}
%
%
%
%
%
%
%
%
%
	\begin{tikzpicture}
	\foreach \x in {0,0.2,...,{3.8}} {
		\draw (\x,1.5)--(\x+0.15,1.5+0.15);
	}
	\draw (0,1.5) -- (4,1.5);
	\foreach \x in {0,0.2,...,3.8} {
		\draw (\x,0-0.15)--(\x+0.15,0);
	}
	\draw (0,0)--(4,0);
	
	\draw (3.2,0.05) rectangle +(0.1,1.4);
	\draw[->] (3.2,1.5/2)-- +(-0.4,0);
	
	\particle{(1.2,0.8)}{0.08}{0.2}{-12}
	\particle{(1.81,0.47)}{0.16}{0.15}{-53}
	\particle{(1.92,1.12)}{0.31}{0.4}{15}
	\particle{(2.72,0.3)}{0.2}{0.2}{-20}
	
	\end{tikzpicture}
	\caption{In the infinite tube, quasi-one-dimensionality can be achieved by inserting a movable `piston' preventing dispersal of the particles. }\label{fig:pistons}
\end{figure}
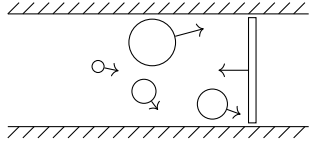

For the infinite tube with a finite number of particles to have a sensible long-time limit, the particle's properties must be such that all particles form a cluster with a shared group velocity. Either, the particles may be effectively single-file, or they can overtake but we add a movable piston---with its own velocity and diffusivity---as shown in \autoref{fig:pistons}.

Suppose that there is a a phase space velocity $\vu$ such that
\begin{equation}
P(\vx,t) \overset{t\to\infty}{\longrightarrow} P'(\vx') = P'(\vx - \vu t) 
\end{equation}
Then the FPE \eqref{eq:FPE} implies
\begin{equation}
\vec{J}(\vx,t) \overset{t\to\infty}{\longrightarrow}   \vec{J}'(\vec{x}') =  \vu P'(\vx').
\end{equation}
Then we can apply most of the previous section done for the periodic tube to $P', \vec{J}'$, with the conditions---instead of periodicity---that $P'$ must be normalizable over the now infinite domain $\Gamma$: 
\begin{equation}
\int_\Gamma d\vx\, \exp[ \vv^\top\mathsf{H} \vx] < \infty.
\end{equation}
The precise conditions on the drift velocities and diffusivities will depend on whether the particles are forced into single-file or not.

\section{Discussion}\label{sec:disc}

We have used a minimalistic model of hard spheres, drifting and diffusing in a periodic tube, to show how heterogeneity in particle properties together with volume exclusion interactions determine the global current in a non-trivial way. Specifically, the slower-diffusing particles have a greater influence on the current, which is explained through an inter-particle ratchet effect---one to be added to a growing catalogue of ratchet effects in non-equilibrium systems. Our technical results rest upon an assumption of quasi-one-dimensionality in an extended sense---the inability of any particle to overtake all others, or alternatively the presence of moving walls, such that all particles are geometrically constrained to a common net velocity in the steady state. We studied several strictly one-dimension scenarios. Then the  general-dimensional case was treated, and a formal explanation for the solvability of the model under quasi-one-dimensionality and tube homogeneity was provided: a unique direction in phase space exists, which never encounters a boundary, is in fact always perpendicular to the all boundaries, and the system is invariant in this direction. This direction is that of the irreversible drift, whose magnitude arises  microscopically through particle interaction under non-equilibrium conditions, but which affords a change of reference frame relative to which the system obeys detailed balance.

The global current, expressed by Eq.\ \eqref{eq:intro-nu} or its one-dimensional version Eq.\ \eqref{eq:SFD-nu-1}, showcases the principle that under non-equilibrium conditions, time-symmetric parameters can have an arbitrarily large influence even on time-asymmetric observables \cite{Maes2018}. Here, diffusivities are the time-symmetric quantities, which have a non-linear and unrestrained impact on the particle current in our model. In contrast, the steady state and linear response of an equilibrium system is entirely determined by the dissipative model parameters. As noted at the end of the \hyperlink{foo}{Introduction}, because of the way the whole diffusion matrix $\mathsf{D}$ enters the expression for the current, non-diagonality, arising from noise-correlations between particles or spatial bias in the diffusion, may be a source of new non-dissipative effects on the current. The generalization of the ratchet effect to a non-diagonal diffusion matrix would be that the current is most influenced by the eigenmode of diffusion with smallest eigenvalue. We invite the identification of a physically motivated problem where this generalized ratchet effect is observed. 

Other modifications to the minimalistic tube model may deserve consideration. For instance, what is the impact of a non-spherical particle shape on the described ratchet effect? The problem with incorporating this feature in the mathematical framework we have employed, is not so much the extra complication to the reflective boundary conditions, but that degrees of freedom describing the orientation of a particle and their dynamic must be defined. The integrability conditions derived from the irreversible drift would almost certainly impose severe limitations on which coupling between orientation and other degrees of freedom would be allowed.

Beyond deriving the integrability conditions, we have not explicitly investigated any particle interactions beyond volume exclusion. In Ref.\ \cite{Wood2019}, spring forces between the filaments ratcheting a membrane could be analysed, and presumably relevant interactions could also be treated for other special cases of the tube model, with the current then governed by Eq.\ \eqref{eq:u(x)}.

A different direction would be to study the open-boundary tube, where particles enter with properties drawn from some distribution. One might anticipate different phases of the current as injection rates are altered, as well as different phase diagrams depending on the distribution of properties chosen, e.g.\ with regards to the weight of the tails in the reciprocal diffusivity distribution.

In conclusion, particle heterogeneity under nonequilibrium conditions is likely to be a source of many collective non-dissipative phenomena, with inter-particle ratcheting being but one exemplar.  

\addcontentsline{toc}{section}{Acknowledgments}
\section*{Acknowledgments}

Emil Mallmin acknowledges studentship funding from EPSRC Grant No.\ EP/N509644/1.

\addcontentsline{toc}{section}{References}
\section*{References}

\bibliographystyle{physicsbibstyle}	
\bibliography{MasterRef_copy}

\end{document}